%% file: sosym2019.tex
\RequirePackage{fix-cm}
\documentclass[smallextended]{svjour3}

\usepackage{makeidx}
\usepackage{graphicx}
\usepackage{url}
\usepackage{cite}
\usepackage{adjustbox}
\usepackage{float}
\usepackage{rotfloat}
\usepackage{rotating}
\usepackage{subcaption}
\usepackage{multirow,booktabs}
\usepackage{tabularx}
\usepackage{diagbox}
\usepackage{multicol}
\usepackage{booktabs}
\usepackage{amssymb}
\usepackage{pifont}

\newcommand{\nameprog}[1]{\texttt{#1}}
\newcommand{\cmark}{\ding{51}}%
\newcommand{\xmark}{\ding{55}}%

\begin{document}

\title{Model-Driven Process Enactment for NFV Systems with MAPLE}

\titlerunning{Process Enactment for NFV Systems with MAPLE} 

\author{Sadaf Mustafiz \and Omar Hassane \and Guillaume Dupont \and Ferhat Khendek \and Maria Toeroe}
\authorrunning{S. Mustafiz, O. Hassane, G. Dupont, F. Khendek, and M. Toeroe}   

\institute{S. Mustafiz \at SCS, Ryerson University, Toronto, ON, Canada\\
	\email{sadaf.mustafiz@ryerson.ca}
	\and
	O. Hassane \at ECE, Concordia University, Montreal, QC, Canada\\
	\email{o\_assane@encs.concordia.ca}
	\and
	G. Dupont \at ECE, Concordia University, Montreal, QC, Canada\\
	\email{gdupont@encs.concordia.ca}
	\and
	F. Khendek \at ECE, Concordia University, Montreal, QC, Canada\\
	\email{ferhat.khendek@concordia.ca}
	\and
	M. Toeroe \at Ericsson Inc., Montreal, QC, Canada\\
	\email{maria.toeroe@ericsson.com}
}

\date{}

\maketitle

\begin{abstract}
	
	The Network Functions Virtualization (NFV) advent is making way for the rapid deployment of network services (NS) for telecoms. Automation of network service management is one of the main challenges currently faced by the NFV community.
	Explicitly defining a process for the design, deployment, and management of network services and automating it is therefore highly desirable and beneficial for NFV systems. The use of model-driven orchestration means has been advocated in this context. As part of this effort to support automated process execution, we propose a process enactment approach with NFV systems as the target application domain. Our process enactment approach is megamodel-based. An integrated process modelling and enactment environment, MAPLE, has been built into Papyrus for this purpose. Process modelling is carried out with UML activity diagrams. The enactment environment transforms the process model to a model transformation chain, and then orchestrates it with the use of megamodels. In this paper we present our approach and environment MAPLE, its recent extension with new features as well as application to an enriched case study consisting of NS design and onboarding process.

	\keywords{Process Enactment \and Megamodelling \and Papyrus \and Network Functions Virtualization (NFV)}

\end{abstract}

\section{Introduction}
\input{introduction.tex}

\section{Background}
\input{background.tex}

\section{Process Enactment Approach}
\input{environment.tex}

\section{Tool Support}
\input{architecture.tex}

\section{NFV Case Study}
\input{casestudy.tex}

\section{Related Work}
\input{relatedwork-rev.tex}

\section{Conclusion}
\input{conclusion.tex}

\begin{acknowledgements}
This work is partly funded by NSERC and Ericsson, and carried out within MAGIC, the NSERC/Ericsson Industrial Research Chair in Model Based Software Management. 
\end{acknowledgements}

\bibliographystyle{plain}
\bibliography{bibliography}

\end{document}

%% file: introduction.tex
Automating the end-to-end management of network services (NS), in other words, enacting the workflow or process for network service management without manual intervention is highly desirable in the telecommunications domain and remains a major challenge for network operators and service providers~\cite{mijumbi2016management, chen2015realizing}. Network Function Virtualization (NFV) builds on cloud computing and has the ultimate goal of automating provisioning and management of network services - an essential feature for 5G systems. The European Telecom Standards Institute (ETSI) has recently launched a zero-touch network and service management group.
As stated in \cite{zsm-dec2017}, the challenges of 5G will trigger the need for a radical change in the way networks and services are managed and orchestrated. 

We believe the application of model-driven engineering (MDE) methods and tools is essential to further such developments in the NFV domain~\cite{berezin-cloudify2017, mde-magic-onap-2017}. 
MDE advocates the use of models as first class citizens in the engineering process. The models are manipulated via transformations which form the backbone for automation in MDE.
ETSI has recently released an information model for NFV~\cite{etsi2017mano-im}. Leveraging these models can substantially benefit the NFV systems by reducing their development and management efforts. Moreover, explicit modelling of the process not only allows for the automation of the NS management process but also paves the way for streamlining and optimization. Such a process model (PM) can potentially be mapped to model transformation chains hence enabling NS management and orchestration via model-driven process enactment\cite{wires-mtatl2009, etien-amt2012, basciani-models2014}. 

Previously, we have proposed a model-based process for NS design and deployment~\cite{mptk16_nca}. The proposed workflow is compliant with the NFV reference framework, and is a first step towards the necessary automation of the NS design and deployment process for NFV systems. We followed up the work in \cite{sdl2017} by elaborating on a method for NS design and proposing an initial approach for enacting the NS design, deployment and management process. 
In this paper, we focus on the enactment approach and present an integrated process enactment environment for NFV systems. MAPLE (\textbf{MA}GIC \textbf{P}rocess Mode\textbf{l}ling and \textbf{E}nactment Environment) provides support for model management with the use of megamodels. We demonstrate the use of MAPLE on the NS design and onboarding process, which represents a portion of the NS management process.   
We adapt the Papyrus~\cite{papyrus} environment to provide tool support for process enactment. Papyrus is the tool of choice of ETSI NFV. 

This paper is an extension of \cite{maple-ecmfa2018} and focuses on addressing the following challenges in process enactment: enactment of PMs with hierarchy, and enactment of heterogeneous (cross-technology) transformation chains. It introduces new features in MAPLE along with an extended and more complex NFV application. We have added support to allow enactment of a PM with multiple activities as well as enactment of a heterogeneous model transformation chain. We have integrated a Java handler which enables MAPLE to detect and execute Java transformations in addition to ATL (ATLAS Transformation Language) transformations. We have also incorporated a \emph{process module} for running executable processes as part of a PM enactment. The megamodel is now updated on-the-fly during enactment with the new generated instances. 
We have extended the NFV case study used as the target application by adding the NS onboarding part. A demo video covering the new features and the extended case study has been made available. In addition, we provide more details on the approach and the backend, and also expanded the related work section.

The rest of the paper is structured as follows: Section~\ref{sec:background} gives a brief background on process modelling, megamodelling, and transformation chaining. Section~\ref{sec:environment} presents the enactment approach. Section~\ref{sec:toolsupport} discusses the tool support and the MAPLE architecture.
Section~\ref{sec:casestudy} presents our NFV case study, and demonstrates the use of the environment on the NS design and onboarding process. 
Section~\ref{sec:relatedwork} discusses and compares related work. Finally, Section~\ref{sec:conclusion} concludes with some future work.

%% file: background.tex
\label{sec:background}

This section provides a brief background on some of the underlying concepts, namely process models, megamodels, and transformation chains.

\subsection{Process Models} 
In our work, we use UML 2.0 Activity Diagrams~\cite{umlspec} to represent and visualize a process model. Activity Diagrams are typically used to model software and business processes. They allow the modelling of concurrent processes and their synchronization with the use of fork and join nodes. Both control-flow and object-flow can be depicted in the model. An activity node can either be a simple \emph{action} (representing a single step within an activity) or an \emph{activity} (representing a decomposable activity which embeds actions and/or other activities). An activity specifies a behaviour that may be reused, i.e., an activity can be included in other activity diagrams to invoke behaviour.
Along with the activities, the input and output models associated with each activity are also clearly identified via input and output parameter nodes (denoted by rectangles on the activity border). 
Since UML 2.0 Activity Diagrams are given semantics in terms of Petri Nets~\cite{umlspec}, the precise formal semantics allow the activity diagrams to be simulated and analyzed. 

We use Papyrus~\cite{papyrus} to create process models. Papyrus is an open-source Eclipse-based UML 2 modelling environment which mainly provides a graphical editor for creating UML 2 compliant diagrams, but also includes extensive support for SysML, UML-RT as well as other UML-based domain-specific languages.
One of the core ideas of Papyrus is that it is completely customizable: appearance, diagrams, palettes, etc. Everything (in theory) can be tweaked, allowing one to use this tool as a base for building custom environments fairly easily.
Nowadays, Papyrus is widely used in industry, and it is thus far more interesting and useful to develop plug-ins for this platform than to create standalone, isolated programs.

\subsection{Megamodels} 
Model management approaches typically use megamodels which provide structures to avoid the so-called `meta-muddle'~\cite{bezivin2005}. The philosophy behind megamodels is that \textit{everything is a model} - models themselves, of course, but also metamodels, transformations and even other resources - and models can be linked together through relations (such as, conformance or derivation).
A megamodel contains artifacts (which are models), relations between them (which may be transformations), and other relevant metadata. 
It can be seen as a map to find and link together all involved models. It forms a repository of models, transformations, and even tools. 

It also can be used to enforce conformance and compatibility checks between the various models and transformations. It is also useful for reusing and composing transformations in transformation chains. 

\subsection{Transformation Chains}
Model transformations (MT) can be composed, typically in a pipeline architecture, as a model transformation composition - generally, referred to as a model transformation chain~\cite{czarnecki2002}. In such a composition, the output of a transformation becomes the input of a subsequent transformation and so on. This results in a chain of model-to-model and/or model-to-text transformations. Such chains enable integration of transformations developed in multiple languages~\cite{garces2017phd}. 

Transformation chaining is the preferred technique for modelling the orchestration of different model transformations~\cite{mdse2017}. Orchestration languages are used for the composition of the transformations in order to model the chain as sequential steps of transformations. Complex chains can incorporate conditional branches and loops, and can also model composite chains. 

Such transformation chains are very useful for decomposing a process into simpler modules (divide and conquer) in order to provide better maintainability, reusability, and extensibility~\cite{mdse2017}. Similarly, MT chains can also be used to build complex transformations based on existing transformations. Textual languages, such as ANT\cite{ant}, are often used to define MT chains.

%% file: environment.tex
\label{sec:environment}

\begin{figure}[tb]
	\centering
	\includegraphics[width=1\columnwidth]{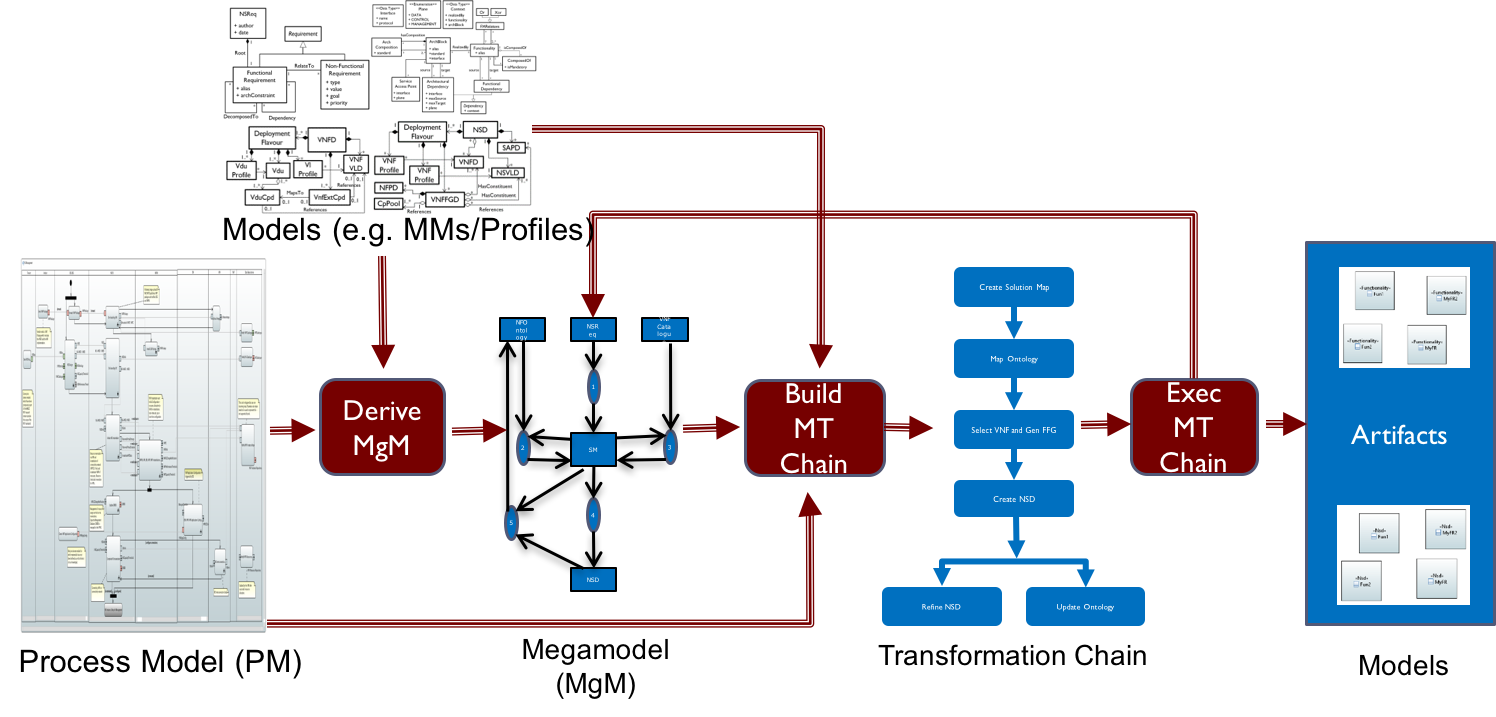}
	\caption{Process enactment approach}
	\label{fig:approach}
\end{figure}

In MAPLE, process enactment is carried out with the use of transformation chain orchestration in combination with model management means.
Figure~\ref{fig:approach} gives an overview of our enactment approach.

As mentioned earlier, in our work we use UML Activity Diagrams (AD) to represent and visualize Process Models. ADs are typically used to model software and business processes. It is suitable for modelling dynamic system behaviour and for capturing software design and architecture-level details. Process models created using such a formalism has the added advantage of enabling synchronization with other UML architectural models. It is also possible for such processes to be modelled with some other workflow modelling language, for instance BPMN\cite{bpmn2011}. 

The Eclipse Papyrus Activity Diagram environment is used to create a PM. The PM instance conforms to the UML 2.0 Activity Diagram language. Each PM includes a set of activities which can be defined with one or more actions. In our work, the behaviour of each of these activities is typically implemented with a set of model transformations. However, it is also possible that some of the activities/actions reuse existing source code (for instance, written in Java, C, or Python). 
For our purpose, we need to associate these actions with the model transformations or the executables which implement them. We use the attributes of the activity nodes in the diagram to add information about the associated transformations. 

Each action in the PM is also associated with a set of input and output models. Papyrus requires all metamodels to be mapped to profiles to allow model instances to be created and to be used as source or target models of the transformations. As per the ETSI NFV modelling guidelines, our models also adhere to the NFV Papyrus \emph{OpenModelProfile}~\cite{etsi2017-im-papyrus}.

\subsection{Deriving the Megamodel (MgM)}
One of the main issues we address in this work is related to resource management: how can we centralize information about resources we need so that we can easily access them? The problem induced by this question is actually not trivial: a transformation involves several metamodels that can be expressed using heterogeneous technologies. The transformation itself can be considered as a model conforming to a specific metamodel, for instance ATL~\cite{atl_models2006}, QVT~\cite{qvt2011}, Epsilon\cite{epsilon2008}, etc. Besides, transformations define a very precise configuration as to what must be fed in and produced, something we need to be aware of whenever we run the transformations. To effectively handle the multitude of models that we are dealing with in our approach, we employ the use of megamodels for model management in MAPLE. 

An MgM is derived in two steps: 1) by registering the resources, and then 2) by registering the PM and weaving the PM/MgM. 

\paragraph{Registering the resources.}
To begin with, the resources which are part of the project (metamodels/profiles) are registered in the MgM. This is carried out automatically by going through the project workspace (referred to as \emph{workspace discovery}), and an initial MgM is derived at this stage.
 
\paragraph{Registering the PM.}
Following the workspace discovery, the MgM is incrementally built by carrying out a \emph{PM discovery}. This step involves registering the PM and the associated model transformations and executables in the MgM. 

\paragraph{Weaving the PM and the MgM.}
The PM needs to be linked with the elements of the MgM so that we can reach, when needed, every information relevant to enact it. However, we do not want any constraint on the shape of the PM, effectively decoupling its metamodel as much as possible from the MgM metamodel.
Therefore, the MgM should be ideally independent from the PM, since it is meant for keeping track of resources. We do not want to (and actually cannot) refer to the MgM in the PM, but we still need a link between these two entities.
This link is created by weaving the PM and the MgM, and storing the details in a \emph{weave model}. A weave model is a special kind of model that defines relationships between the objects and relations of other distinct models (at least two)~\cite{weave1,jouault2010}. 
The weave model binds every relevant element of the PM to their corresponding resources in the MgM, without touching the structure of either of them.
The weave model is dependent on the PM. In other words, the weave metamodel is always specific to the PM language it weaves. 
Thus, if we had to adapt the environment for another PM language (e.g., expressed with BPMN), it would be necessary to create a new weaver to bind the new type of PM with the MgM. 

The PM representation (modelling language that it conforms to) is only known following the PM registration step, hence the weaver to use can only be identified at this point. Since the weaving is carried out after the registration of the PM, the weave model and its metamodel are only added to the MgM later and does not show up in the base MgM. This supports out goal was to make the environment generic and extensible for other potential PM representations. The weave metamodel currently supported is specific to a subset of UML Activity Diagrams (i.e. it weaves elements of the UML AD to elements of the MgM). 

\subsection{Building the Transformation Chain}
The PM is given translational semantics by mapping it to a transformation chain. The chain is in essence a schedule with the required details (sequence of actions, transformations used, inputs and outputs of the transformations). 
This allows us to build a generic enacter, instead of having an enacter for each kind of PM. Having a generic enacter also leaves scope for integrating other formalisms for modelling the PM. 

The translation from a PM to a transformation chain is implemented as an ATL model transformation, which takes as input various data (the PM, the weave model, the MgM and if applicable, additional environment information) and yields the corresponding transformation chain.

\subsection{Executing the Transformation Chain}

\paragraph{Executing the Chain.} Once the transformation chain is created, we need to be able to execute the chain in order to enact the PM. 
For this purpose, we developed an \textit{enacter}, which is simply a program that can execute the correct actions in the right order, based on a schedule, namely the transformation chain model.

Similar to UML 2 Activity Diagrams, the generated chain is also given token-based semantics.
Therefore, the enacter developed is based on controlling the tokens and activating the actions when needed.
Concurrent access to the same model instance is resolved at this stage to avoid model inconsistencies.

\paragraph{Updating the Megamodel (MgM).} The MgM is first derived based on existing resources (prior to enactment). During enactment, the MgM gets dynamically updated with new resources. With each model transformation execution, the generated artifacts are added to the MgM. The MgM is updated with a reference to each new model instance. The conformance links to the metamodels for the new instances are also retained in the MgM.

%% file: architecture.tex
\label{sec:toolsupport}

This section covers the support provided by the MAPLE environment and the backend architecture of MAPLE.
\subsection{Language Support}
\label{subsec:langsupport}

With regards to the language support  in MAPLE, our main goal was to have an extensible cross-platform environment, which would allow any technology, transformation type/engine, and PM language to be adapted and accommodated at a later time. 

A process model can be composed of one to many activities running in sequence or in parallel. The activities themselves can be composed of actions and/or sub-activities. Hence, a PM can have nested PMs. Initially MAPLE allowed the enactment of a PM without hierarchy only. The enactment was also only possible of a PM in which actions were implemented as ATL transformations. To enact more complex PMs, we needed MAPLE to support the notion of hierarchy and also execution of actions implemented in different transformation languages. 

Recently, support for Java was also integrated in MAPLE. This allows the behaviour of activities in the PM to be implemented in Java. 
In order to support implementations written in various languages in MAPLE, we have come up with a programming language agnostic module that provides support for enacting PMs which include executable actions, i.e., actions associated with operating system processes. 
Similar to the support for ATL and Java transformations, the executable code associated with an action needs to be identified as an attribute of the action in the Papyrus Activity Diagram environment. The specification of the language is optional. The details including the language, input parameters, and output parameters need to be defined in a separate configuration file. The base megamodel has been extended to recognize these executable processes. Currently, all executables conform to a base metamodel named \texttt{process} in the megamodel.

We have extended MAPLE with a built-in module, referred to as the \textit{process executable} module which allows transformations written in any specialized model transformation language (e.g., Epsilon, QVT, Kermata) or general-purpose programming language (e.g., Java, Python, C) to be enacted as part of the PM. The \textit{executable process} is created as a file having the ``.process'' extension. Its main component is the \textit{process} element which has a command and a name attribute. The command attribute takes as input the OS command to launch, and the name attribute which represents the name of the \textit{executable process}. Each \textit{process} element in the process module file can contain parameters which represent the command arguments. 

\begin{figure}[tb]
	\begin{minipage}{0.5\linewidth}
		\centering
		\includegraphics[width=0.8\textwidth]{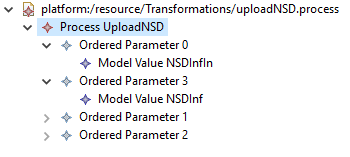}
		\vspace{-5mm}
		\caption{A sample .process executable file}
		\label{fig:processFile}
	\end{minipage}%
	\begin{minipage}{0.5\linewidth}
		\centering
	\includegraphics[width=1\textwidth]{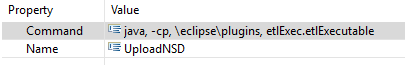}
	\caption{Process element properties in a process executable file}
	\label{fig:commandProperty}
	\end{minipage}
\end{figure}

\begin{figure}[h]
	\centering
		\includegraphics[width=0.6\textwidth]{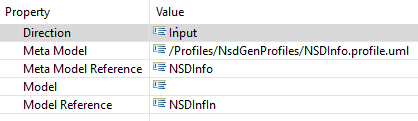}
		\caption{Process parameters properties in a process executable file}
		\label{fig:processParameters}
\end{figure}

In order to run the \textit{executable process} file, the information including the OS command, input parameters, and output parameters (which can be of type UML or of any other type, such as Ecore or XMI) need to be specified. Fig.~\ref{fig:processFile}, shows a sample \textit{executable process} file in which the root process element contains input and output parameters. The root process element defines the command to execute as shown in Fig.~\ref{fig:commandProperty}, as for the parameters (shown in Fig.~\ref{fig:processParameters}), they define the direction of the model, i.e., whether it is an input or an output, its metamodel and a model reference that matches the corresponding model reference in the PM.

While this extension makes it easy to incorporate executable files and use it within the PM, it should be noted that in this case no information is retained in the MgM regarding the transformation language used.

\subsection{Architecture}
\label{sec:architecture}

\begin{figure}[tb]
	\centering
	\includegraphics[width=0.9\textwidth] {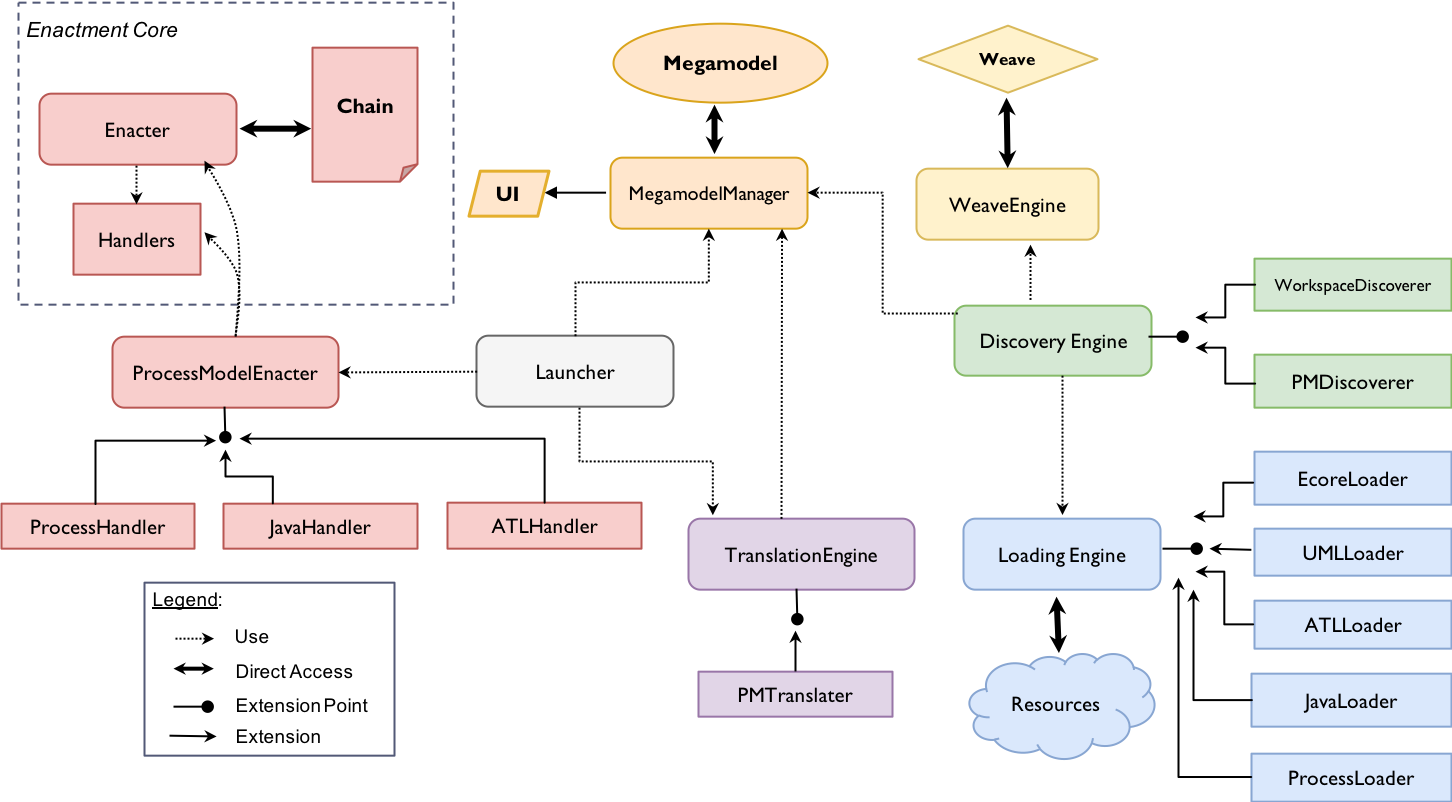}
	\caption{Backend architecture}
	\label{fig:backend}
\end{figure}

The backend architecture of the enactment extension developed for Papyrus (Neon Release) is shown in Fig.~\ref{fig:backend}. In the figure, the core functionalities are represented using rountangles (rounded-corner rectangles). The optional functionalities (extensions) are represented by rectangles. The architecture is open for future extensions. 

\paragraph{Model Loading.}
This part corresponds to the blue boxes in Fig.~\ref{fig:backend} and is centered around the \textit{loading engine}. The goal of this subsystem is to provide a high level interface with the actual resources of the system (i.e., the file system). It is able to recognize the correct way of loading a file, and to extract from the file data, which could be interesting to have in the megamodel, for example. Typically, when trying to register a file in the MgM, the discovery engine asks the loading engine to load the file to be able to extract data. Defining \textit{loaders} (such as, \nameprog{EcoreLoader}, \nameprog{UMLLoader}, \nameprog{ATLLoader}, \nameprog{JavaLoader}) allows for different technologies to be incorporated in the global system, making it able to understand new formats. In the case of Java, the \nameprog{Java} loader parses Java programs which conform to the EMF Java metamodel\footnote{https://download.eclipse.org/modeling/emf/emf/javadoc/2.6.0/org/eclipse/emf/java/package-summary.html}. 

To address \textit{executable process}es, we defined an \textit{executable process} loader responsible for reading related configuration files and extracting useful data from them (such as the command to be executed, the references of the metamodels to which the in/out instances conform to, etc). These executables are seen as resources conforming to a base metamodel named \texttt{process} in the MgM. 

The MAPLE backend is designed such that it is modular and easily extensible. This makes it possible to add new loaders without ``touching'' the core loading engine or the existing loaders. 
Note that one has to extend the base MgM with the metamodels of the new loaders.

\paragraph{Discovery.}
\label{sssec:registering}
This is centered around the \textit{discovery engine} and corresponds to the green boxes in Fig.~\ref{fig:backend}. The \emph{discovery} is the process by which a resource is added to the MgM. It includes the recognition of the resource (determining that it is a metamodel, a transformation, a profile, etc.) as well as the extraction of its data (URI, name, inputs and outputs, etc.) with the help of the \emph{loading engine}.  
As in the case of the loaders, it is possible to develop a custom \textit{discoverer} (for instance, as shown in Fig.~\ref{fig:backend}, \nameprog{WorkspaceDiscoverer} or \nameprog{PMDiscoverer}). This allows the tool to be extended with discoverers for other kinds of workflow modelling languages. 

This is also the process that can initiate the \textit{weaving} of the PM and the MgM. 

\paragraph{Weaving the PM and the MgM.} This part corresponds to the yellow boxes in Fig.~\ref{fig:backend}. The \textit{weave engine} creates a weave model containing mappings of several types: ActionMapping (to link the action in the PM and the transformation in the MgM), ObjectNodeMapping (to link the object flow in the PM to the corresponding model in the MgM), and InOutMapping (to link an input/output pin to an object flow to determine the input/output of the transformation it is linked to).

As the weave is entirely dependent on the PM language, it should be noted that the weaving process is completely devolved to the discoverer, which must provide both a specific weaver (an implementation) as well as fill it. In our case, the weave model conforms to a variant of the UML Activity Diagram formalism. 

\paragraph{Megamodel Management.}
This is centered around the \textit{megamodel manager}, the orange part in Fig.~\ref{fig:backend}. It allows the user to access the MgM: request to register a resource, create or delete an MgM, etc. It also includes an extensive API for manipulating the MgM, ensuring its validity throughout the process.

\paragraph{Translation.}
\label{sssec:translation}
This revolves around the \textit{translation engine}, the purple boxes in Fig.~\ref{fig:backend}. The goal is to carry out the translation of a PM to a transformation chain that can be enacted. It exposes an extension point that allows anyone to plug his own translation means, which is required if ever we want to use another type of PM. 
MAPLE supports hierarchy in process models. A process model with multiple activities and/or with nested activities is recognized by the translation engine, flattened and mapped to a transformation chain. The resulting chain includes forks and joins to model concurrent actions.

\paragraph{Scheduling and Enactment.}
This part corresponds to the red part in Fig.~\ref{fig:backend}. It is composed of two subparts: a generic enacter (\nameprog{Enacter}), independent from the project (inside the dashed box labelled ``Enactment Core"), and an interface between this enacter and the remaining part of the project, through the \nameprog{ProcessModelEnacter} part.

It should be noted that the \nameprog{ProcessModelEnacter} is not an independent enacter, but a layer on top of the generic enacter, providing it with specific behaviour as to what to do with each action; that is, actually executing the transformations they correspond to, using interfaces with different transformation engines defined as extensions (e.g.: \nameprog{ATLHandler}, \nameprog{JavaHandler}). For instance, the \nameprog{JavaHandler} compiles and executes the loaded Java source code accordingly. 
A \nameprog{ProcessHandler} which is responsible for launching specific executable actions in the PM has been added to MAPLE. 

\paragraph{Launcher.}
The gray rectangle in Fig.~\ref{fig:backend} is what orchestrates everything needed to execute the PM. It includes the translation to a model transformation chain and the enactment of the chain. Enactment configurations (data associated with the PM to translate and enact it) are generated and stored as a standard Eclipse launch configuration, which are also managed by this part of the tool.

%% file: casestudy.tex
\label{sec:casestudy}

We have used MAPLE to model and enact part of the Network Service Management (NSM) PM proposed in \cite{mptk16_nca}. 
In this section, we demonstrate the enactment of the NS design and onboarding process which are the initial activities of the NS management process. NS management includes activities on a running system hence the execution is time constrained. This NFV case study illustrates the use of MAPLE for enacting a PM composed of multiple activities implemented with a heterogeneous set of transformation languages.

\subsection{Network Service Design and Onboarding}

\paragraph{NS Design.}

A network service (NS), such as VoIP, is a composition of interconnected network function(s) The interconnections explicitly describe the traffic flow between the components.
\emph{Virtualized Network Functions (VNF)} are software pieces representing the resource aspect of network functions in NFV (e.g., a virtual firewall). They are the building blocks of an NS.  
The design of an NS consists of defining an \emph{NS Descriptor (NSD)}, a deployment template which captures the information relevant to NFV. This template is provided to the NFV Orchestrator for the NS lifecycle management.

\begin{figure}[tb!]
	\begin{minipage}{0.5\linewidth}
		\centering
		\includegraphics[width=1\columnwidth]{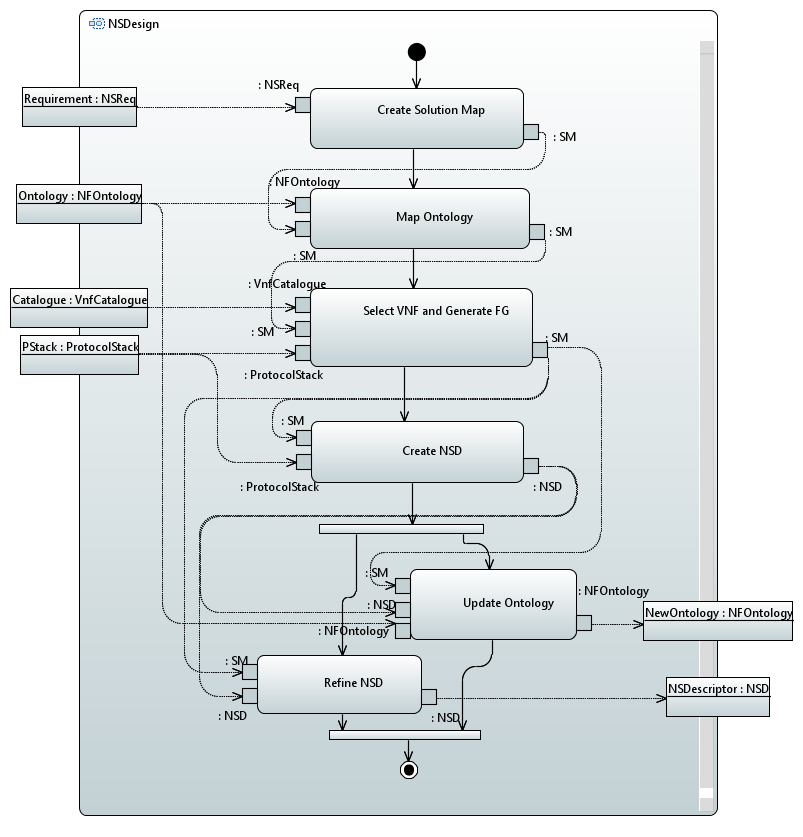}
		\caption{NS Design PM~\cite{sdl2017}}
		\label{fig:nsdesignpm}
	\end{minipage}%
	\begin{minipage}{0.5\linewidth}
		\centering
		\includegraphics[width=1\columnwidth]{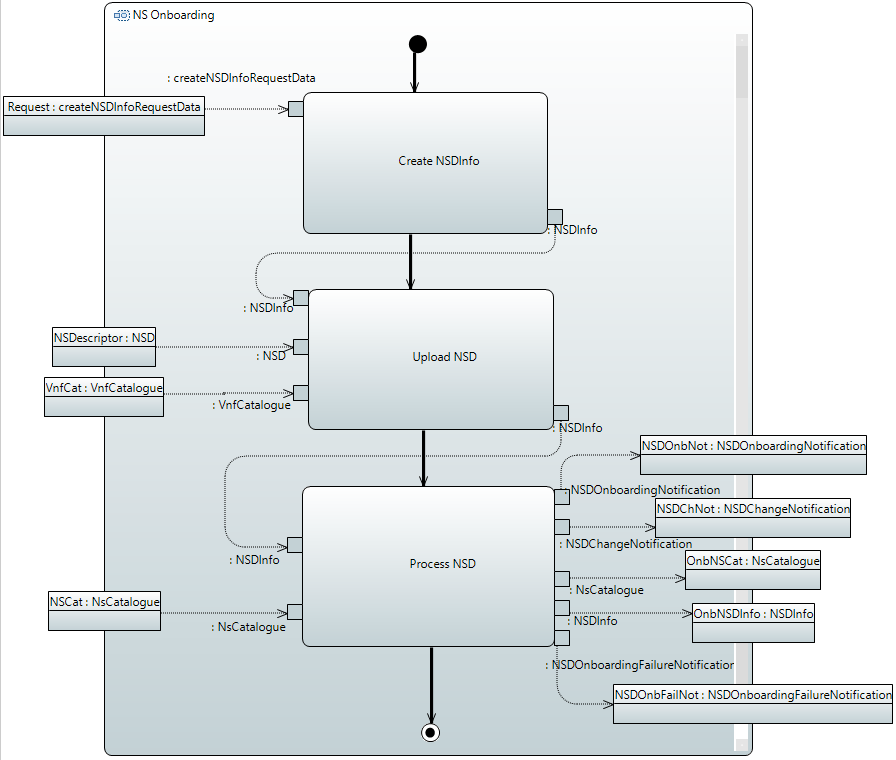}
		\caption{NS Onboarding PM}
		\label{fig:nsonb-pm}
	\end{minipage}
\end{figure}

Coming up with the deployment template for an NS is not an easy task for an inexperienced tenant who has limited knowledge regarding the details of the target NS. Instead of these details, the tenant may request at some level of abstraction the functional and non-functional characteristics of the targeted NS. The gap between these NS requirements (\emph{NSReq}) and the NS deployment template is filled with an automated NS design method. With the help of a network function ontology (\emph{NFOntology}), it is indeed possible to fill this gap and design automatically \emph{NSDs} from \emph{NSReqs}.  
The \emph{NSReq} decomposition is guided by the \emph{NFOntology} to a level where proper network functions can be selected from an existing VNF catalog. After the selection of the VNFs, the method continues with the design of the traffic flows given the characteristics of the selected VNFs and their dependencies. 
These flows are refined further based on the non-functional requirements in the \emph{NSReq}, resulting in the target \emph{NSD}. As a final step, the ontology is enriched with the new decompositions. It is also possible to enrich the ontology with new standards and new services. Please note that the goal of this paper is not to describe the details of the NS design method but to show how the process is enacted using our tool.
The NS design process and the associated modelling languages which are part of the NS Design PM are described in details in \cite{sdl2017}.  
A revised version of the PM is shown in Fig.~\ref{fig:nsdesignpm}.

\paragraph{NS Onboarding.}

This activity is triggered by the NFV Orchestrator due to an incoming \texttt{NSD} (generated by the \textit{NS Design} activity). This step is essential in NFV systems since the NSD needs to be validated to ensure that it conforms to the service provider's platform and then onboarded to the provider's catalog (or repository) of network services. The onboarding activity creates the onboarded NSD artifact (\texttt{NSDInfo}~\cite{etsi2017mano-im}). Once created, the \texttt{NSD} (which is added to the associated \texttt{NSDInfo}) is uploaded and validated. The validation ensures that the VNFs which are part of the deployment template (to be used during instantiation) exist in the provider's VNF catalog. Once the validation passes, the \texttt{NSD} is successfully onboarded and added to the \texttt{NS Catalog}. The target artifacts conform to the ETSI NFV defined information elements. The behaviour of the NS onboarding process is shown in Fig.~\ref{fig:nsonb-pm}.

\subsection{Using MAPLE for NS Design and Onboarding}

\begin{figure}[tb!]
 	\centering
 	\includegraphics[width=1\textwidth]{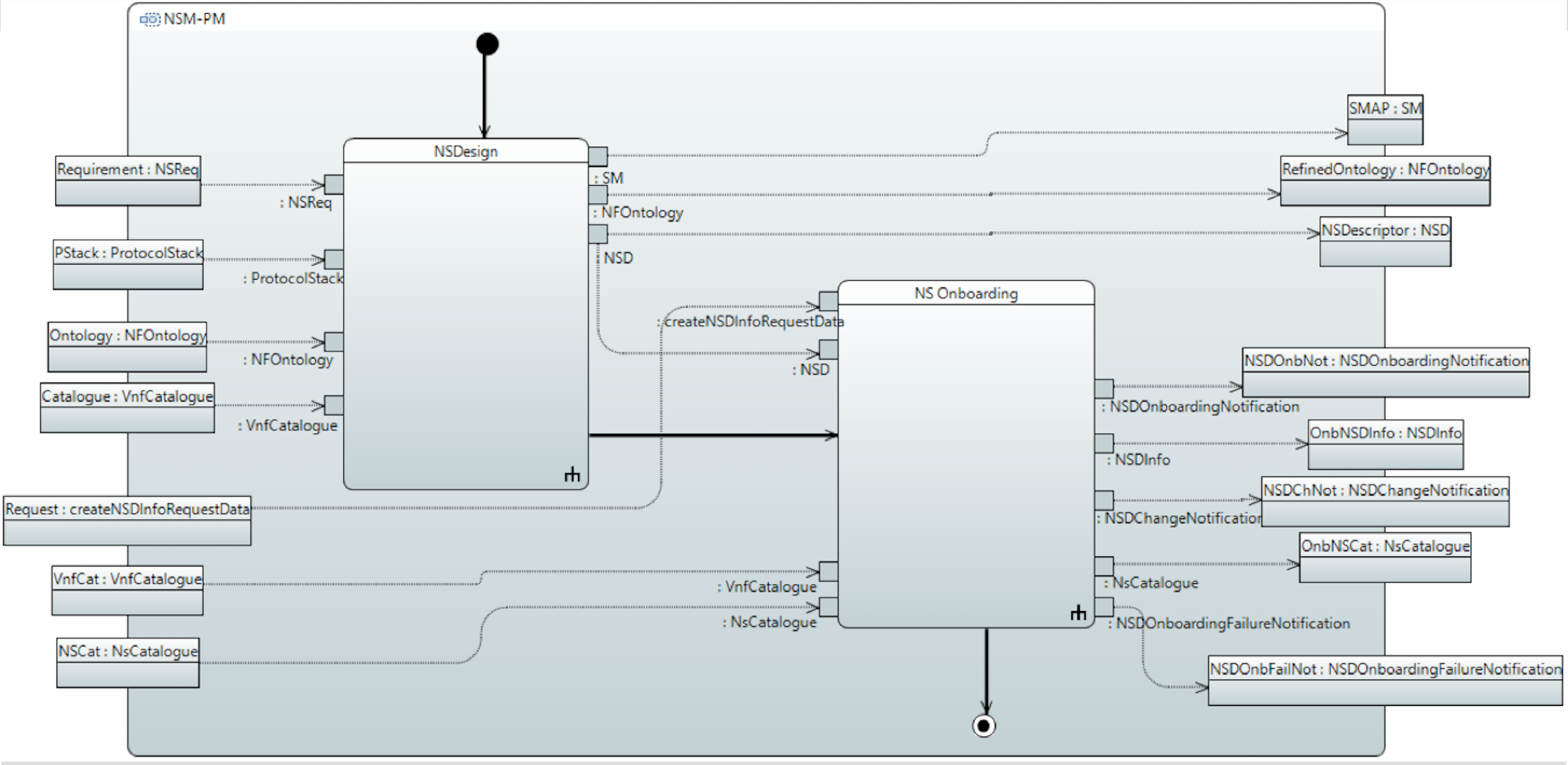}
 	\caption{NSM PM Slice: NS Design and Onboarding PM}
 	\label{fig:nsdopm}
 \end{figure}

 The enactment approach is demonstrated here with a slice of the NSM PM shown in Fig.~\ref{fig:nsdopm}, the NS design and onboarding PM. The PM includes two activities in sequence, the NS design activity followed by the NS onboarding activity, presented in Fig.~\ref{fig:nsdesignpm} and Fig.~\ref{fig:nsonb-pm} respectively. Each of these activities are composed of a set of actions in sequence or in parallel. 
 The enactment process with MAPLE consists of three main steps: deriving the megamodel (MgM), generating the transformation chain corresponding to the PM, and enacting the PM (via the transformation chain) using these resources.
 
 \paragraph{Deriving the Megamodel.}
 
\begin{figure}[tb!]
 	\centering
 	\includegraphics[width=0.4\textwidth]{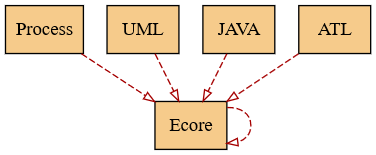}
 	\caption{Base Megamodel (MgM)}
 	\label{fig:initmgm}
 \end{figure}

 Prior to registering any resources, the base MgM initially consists of metamodels of the loaders used to load resources needed for the enactment (see Fig.~\ref{fig:initmgm}). This MgM includes the meta-metamodels (UML and Ecore pre-loaded in the base MgM) and conformance links. In the MgM, the metamodels are represented in orange, UML profiles in green, transformation models in brown, and other models (PM, weave model) in gray. The dashed links represent conformance relationship, and the solid black links represent object flow. 
 
\begin{figure}[tb!]
 	\centering
 	\includegraphics[width=1\textwidth]{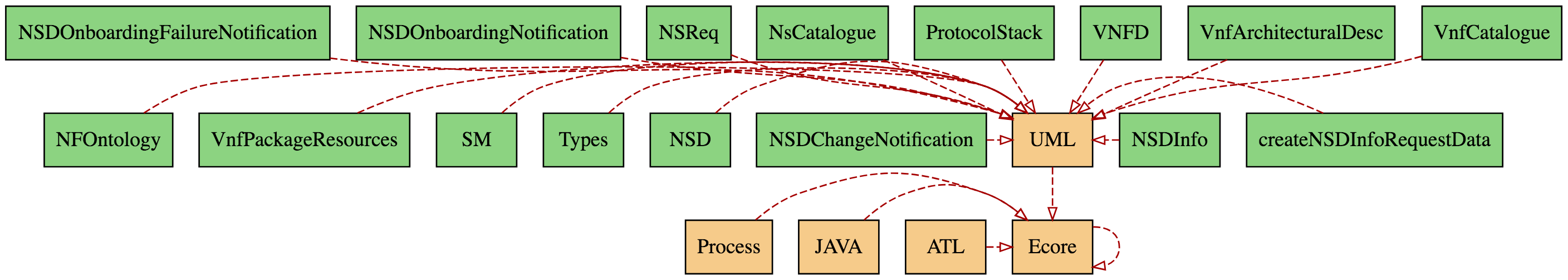}
 	\caption{NS Design and Onboarding: Initial MgM}
 	\label{fig:profilemgm}
 \end{figure}

 We begin by registering the NS design and onboarding resources (profiles for \texttt{NSReq}, \texttt{NFOntology}, etc.) that are used in the process by creating an initial MgM. Once we register the profiles, a conformance relation between the profiles and the UML metamodel is created in the MgM (see graphical view of the MgM in Fig.~\ref{fig:profilemgm}).

 The next step is to register the PM which automatically refines the MgM based on information available in the PM. Each action in the NS Design activity is implemented as an ATL transformation. The NS onboarding is implemented with a heterogeneous set of transformations written in Java, ATL, and Epsilon. Each associated transformation is stored as an attribute of the corresponding activity node. Since there is no handler available for Epsilon transformations in MAPLE, the \emph{executable process} module (discussed in Sections~\ref{subsec:langsupport} and \ref{sec:architecture}) is used to execute the Epsilon transformation. 
 
\begin{figure}[htb!]
 	\centering
 	\includegraphics[width=0.9\textwidth]{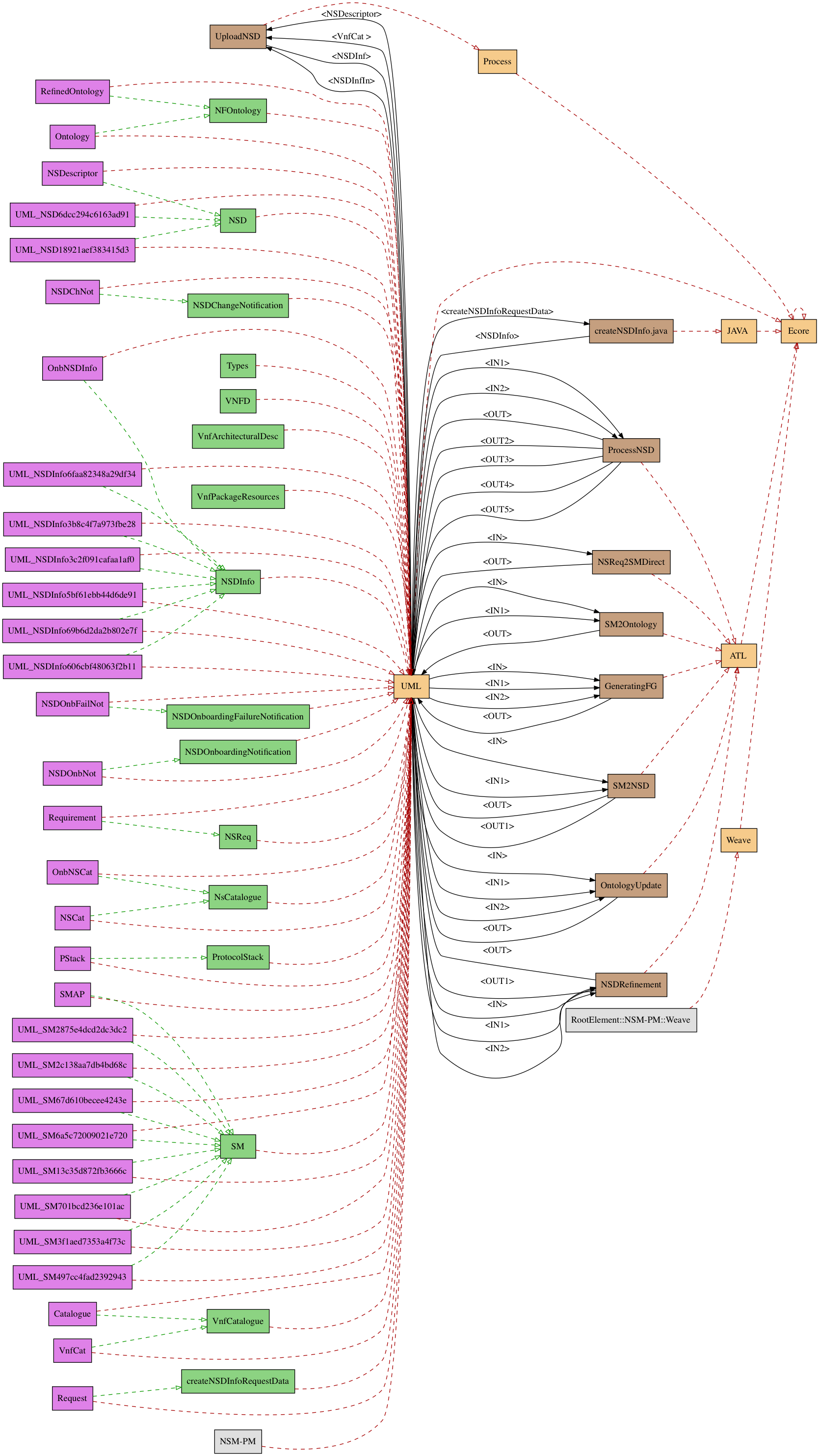}
 	\vspace{-2mm}
 	\caption{NS Design and Onboarding: Updated MgM} 
 	\label{fig:completemgm}
 	\vspace{-10mm}
 \end{figure}

\begin{figure}[tb!]
		\centering
		\includegraphics[width=1\columnwidth]{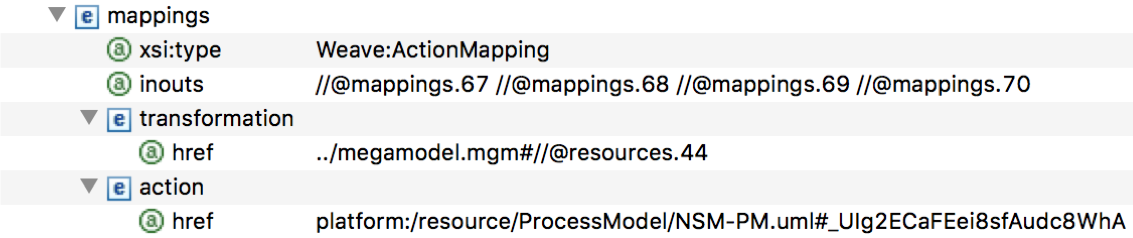}
		\includegraphics[width=1\columnwidth]{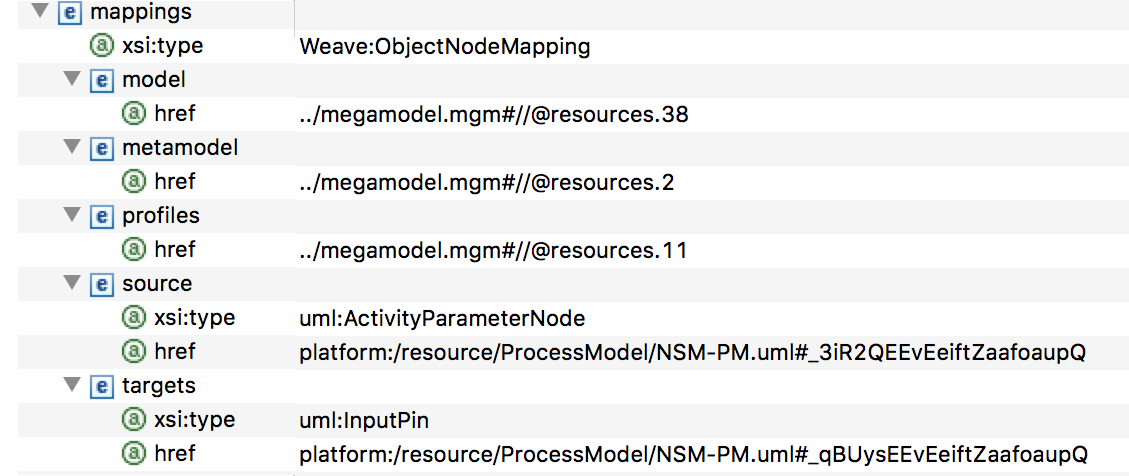}
	\caption{NSM PM Weave Model: Sample Mappings}
	\label{fig:nsdopm-weave}
\end{figure}

 The PM itself is also added as a resource (see Fig.~\ref{fig:completemgm}). While discovering and registering the PM, whenever an object flow links two pins and that flow and pins do not have any assigned name, MAPLE can detect it and create an intermediate model. The weaving of the NS Design and Onboarding PM and the MgM is also carried out at this stage resulting in a weave model containing various mappings. Samples of the mappings part of the generated weave model are shown in Fig.~\ref{fig:nsdopm-weave}.
 
 This main step results in creating an initial repository of models, NFV-specific languages, and tools, along with the relationships between the artifacts. 
 
 \paragraph{Enacting the PM.}
 
\begin{figure}[b!]
	\centering
	\includegraphics[width=0.6\textwidth]{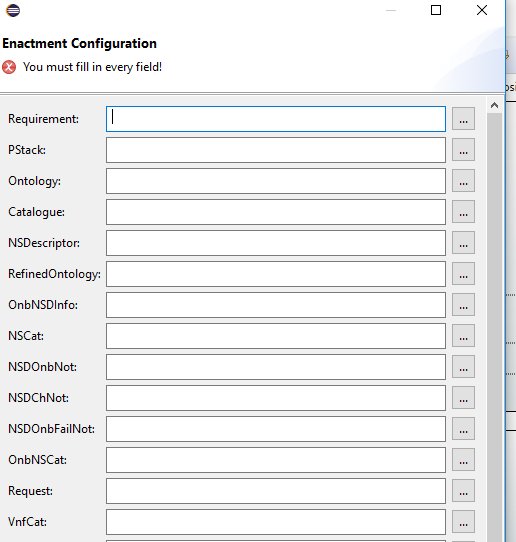}
	\caption{MAPLE Launch Configuration Dialog}
	\label{fig:launchconfig}
\end{figure}

\begin{figure}[tb!]
	\centering
	\includegraphics[width=0.65\textwidth]{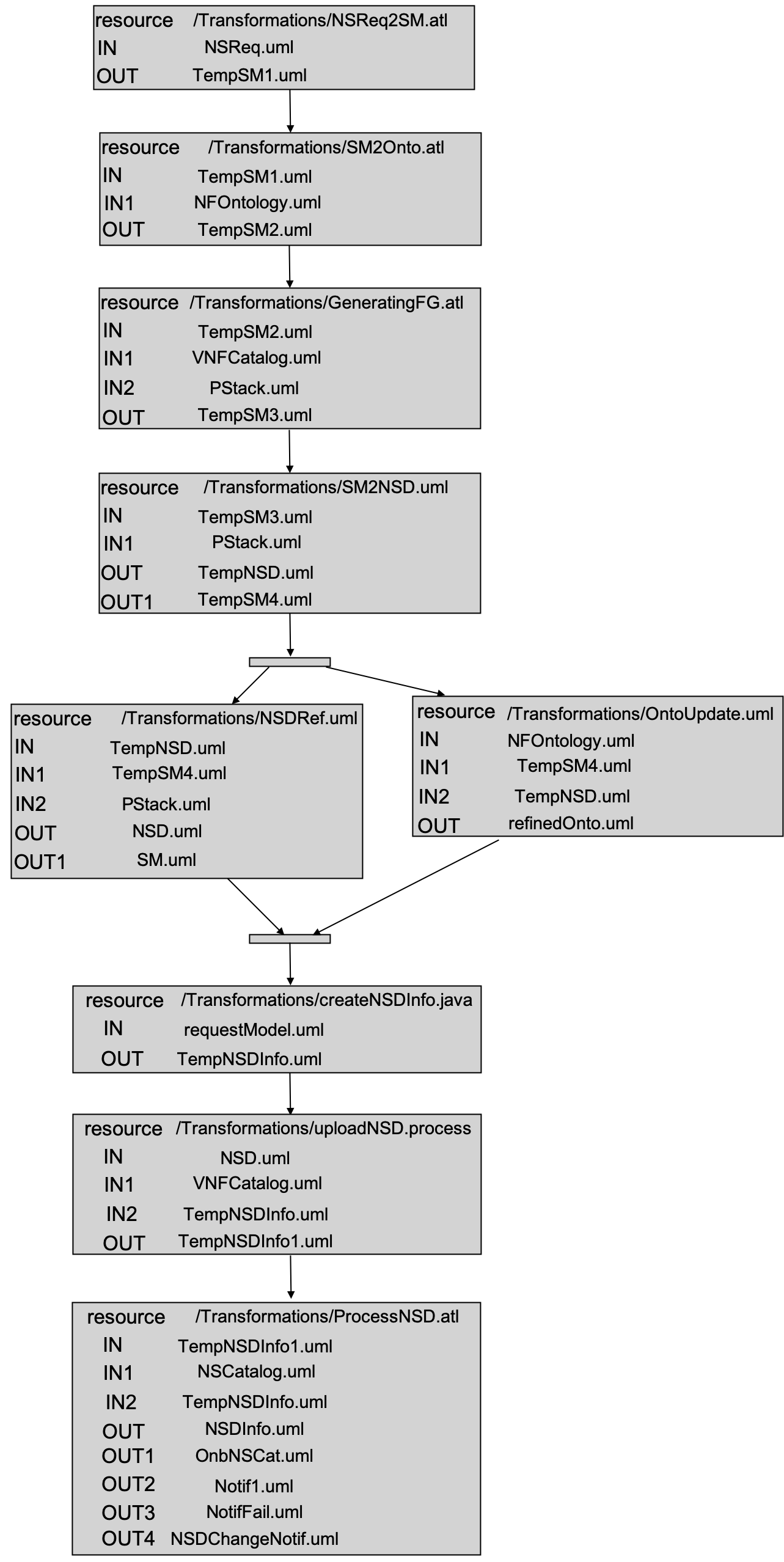}
	\caption{NS Design and Onboarding: Generated Transformation Chain}
	\label{fig:schedule}
\end{figure}

 Once the resources (profiles, transformations and PM) are registered in the MgM, the PM can be enacted. MAPLE based on the PM generates a launch configuration dialog box (see Fig.~\ref{fig:launchconfig}) to fill in the parameters needed for the execution of the PM (such as the different model instances).
 MAPLE can then start the process of enacting by mapping the NS design and onboarding PM to a transformation chain (see Fig.~\ref{fig:schedule}). 
 
Orchestration of the chain is carried out with the use of the embedded orchestration engine. The process execution begins by taking an \texttt{NSReq} model as input and creating an intermediate model, \texttt{SolutionMap}, which is incrementally refined. Once the initial \texttt{NSD} is created, MAPLE allows NSD refinement and ontology enrichment to be carried out concurrently since they are independent of each other, hence optimizing deployment time. The \texttt{ATLHandler} is invoked to execute the actions which lead to the generation of the \texttt{NSD}. The execution of the chain continues with the NS onboarding transformations. MAPLE now invokes the \texttt{JavaHandler} to execute the action to create a skeleton \texttt{NSDInfo}. The enactment continues with the invocation of the \texttt{ProcessHandler} in order to execute the action (implemented in Epsilon) which uploads the \texttt{NSD}. Finally, the \texttt{ATLHandler} is invoked again to execute the next action which involves validating the \texttt{NSD} and updating the catalogs. The enactment ends with the generation or update of the target models, \texttt{NSD},  \texttt{NFOntology}, \texttt{NSDInfo} and the \texttt{NSDOnboarding Notification}s (not shown here). These model instances are dynamically added to the MgM as they are generated.  

With this environment, NFV users with limited modelling expertise and minimal knowledge about the underlying transformations can generate a target NSD with basically a few clicks and onboard it. The configurations for ATL, Java, and Epsilon, ensures that the correct models are passed into each transformation, do not need to be handled by the user. The same PM can be enacted again with different inputs if desired, and it can also be reused as part of another PM if required.

A demo video of the enactment environment is available~\footnote{\url{https://users.encs.concordia.ca/~magic/maple-demo.php}}. The new features of MAPLE are demonstrated in another video~\footnote{\url{https://users.encs.concordia.ca/~magic/maple-demo-new.php}}.

In \cite{maple-ecmfa2018}, MAPLE was only used to enact a single activity which included a set of non-decomposable actions. The NS design and onboarding application demonstrates the use of MAPLE on a larger subset of the PM. This validates that the approach and environment works for 1) PMs with composite activities, 2) PMs implemented with a heterogeneous set of model transformations, and 3) PMs with executable actions.

This work sets the basis for the enactment of the entire NS design, deployment and management process. Each activity in the NS lifecycle involves a complex chain of tasks. We are working on modelling the behaviour of the other activities, e.g. \emph{NS Instantiation} and \emph{VNF Instantiation}. The entire PM can then be mapped on to a composite chain of transformations along with an extended MgM to allow for automated deployment and management of network services. When the NS management PM is enacted using MAPLE, the resulting MgM is a useful repository of NFV-related models, languages, and tools for NFV projects in the industry and academia. 
 
\subsection{Discussion}

In the NFV community, application of MDE is still at the initial stages where information modelling with class diagrams and state machines are more typical. Advanced applications of MDE in this domain is minimal. This work is meant to set the pillars for a project which can advance greatly with the appropriate use of MDE, as well as provide much-needed inspiration to the open-source NFV industry projects to embrace MDE methods and technology. 

The proposed enactment support, MAPLE, has been built into a mainstream, open-source, industry standard tool which is also ETSI NFV's tool of choice - Eclipse Papyrus. MAPLE requires the PM to be created using UML Activity Diagrams in Papyrus. Once the PM is created, enacting it is fairly intuitive using our environment and as elaborated earlier in this section, only requires a few steps on the user end. For the process to be enacted, the actions which are part of an activity need to be implemented. MAPLE allows enactment of PMs with multiple activities composed of actions. The activities can be developed using a single language or multiple languages. Model transformations (implemented with ATL or Java) or executables (Java, C, etc.) can be used for the actions. The latter allows existing code to be integrated and used within the PM. This also enables implementations in other model transformation languages to be handled during the PM enactment. 

MAPLE is easily extensible, and we plan on providing support for other transformation languages including Epsilon and QvT. The tool can also be extended to support other workflow modelling languages. 
One of the main contributions is the underlying support for model management. NFV systems involves numerous heterogeneous and interrelated artifacts which evolve over time. The complexities that arise with the management of such a large and evolving collection of models can be handled with megamodelling techniques. The megamodel forms a repository of models, metamodels, and tools for the NFV domain, which can be of great use in this domain. The generated models are also persisted and dynamically added to the repository to be inspected, compared, or used in transformations at a later time. The MgM makes it possible to explicitly deal with dependencies between models. It is used to carry out type checks between the metamodels of the models to be transformed and to check the compatibility between the metamodels and the transformations by querying the MgM. In case of activities which include a heterogeneous set of transformations, the MgM determines which transformation engine should execute the transformation. 

In addition to being useful for carrying out conformance and compatibility checks, the created MgM can be used for advanced traceability information generation and analysis.
Work is in progress to extend the tool for this purpose. Such a feature would indeed be beneficial for NFV systems.

%% file: relatedwork-rev.tex
\label{sec:relatedwork}

In this section, we first discuss some of the available work on applying MDE in the NFV domain. We then elaborate on how our work compares with the existing approaches and tools for process modelling and enactment in the MDE area.

\subsection{Model-Driven Orchestration Support for NFV Systems}

We have covered the state of the art with regards to NFV management and orchestration in \cite{mptk16_nca}. We only mention relevant model-based or model-driven approaches here.
A few model-based continuous integration and deployment methods and tools have been proposed with cloud as the target domain~\cite{artac2016, ferry2014}, which use domain-specific languages to model the deployment of cloud applications and provide means to adapt the models for use in the runtime environment. Some other model-based approaches exist in the NFV literature~\cite{TOSCA2016, chen2015realizing, lingen2017}. These approaches do not support process modelling.

According to Mijumbi et al.~\cite{mijumbi2016management}, existing NFV industry solutions lack real support for flexibility, interoperability, orchestration and/or automation, which are the core requirements for NFV. Model-driven orchestration has been recently promoted for NFV, and viewed as a more robust method than business process workflows for NFV orchestration~\cite{berezin-cloudify2017}. BPMN-like workflows are in general implementations of specific task-oriented cases which are appropriate for immutable business processes as stated in \cite{berezin-cloudify2017}. In software defined environments~\cite{chiosi2012network} which evolve rapidly, such workflows bring about difficulties and risks.

Recently, several open-source initiatives have been taken by the telecom industry. In particular, the Open Network Automation Platform (ONAP)\footnote{https://www.onap.org} project was formed as an effort to harmonize the telecom industry and provide a common platform for network service orchestration~\cite{onap-wp-2017}. ONAP plans on automating both design and management of network services by deploying model-based methods and tools. A multi-cloud infrastructure orchestration support will be provided with the integration of the Cloudify\footnote{\url{https://cloudify.co}} platform (a TOSCA-based model-driven orchestration platform). Incorporating workflow modelling means is also part of the ONAP roadmap to enable orchestration.

 However till date, the application of advanced MDE techniques, such as model management and transformation chain orchestration, is minimal for NFV systems. With this work, we aimed to show how MDE means can be used in the NFV domain to further the vision of automating end-to-end network service management.

\subsection{Model Management and Process Enactment Support}

Extensive research has been carried out by the model transformations community on composing and executing transformation chains~\cite{etien-amt2012, oldevik-dais2005, wagelaar2006, traco, wires-mtatl2009, aranega-fmtc-2012}. The Model Control Center (MCC) is Eclipse plugin for creating and executing transformation compositions~\cite{mcc2006}. The Juniper project also introduces means for generic transformation chaining~\cite{juniper}. MoTCoF~\cite{motcof2012} is a dedicated model transformation composition framework supporting chaining of black-box transformations. It allows specification of loosely coupled and heterogeneous transformation chains. Similar to our work, some approaches proposed or used a workflow modelling language to define transformation chains at a higher level of abstraction. In \cite{vanhooff2006}, the authors propose a modelling language for defining transformation chains based on a variant of UML Activity Diagrams. This work later progressed into the transformation chaining framework, UNiTi~\cite{uniti2007} (discussed later in this section).
The Wires* framework~\cite{wires-mtatl2009} also adopted the use of UML Activity Diagrams to model transformation compositions. However, UNiTi and Wires* are both data-flow oriented and do not address process-flow.
Tools supporting workflow or transformation chain orchestration also existed at some point~\cite{wires-mtatl2009,uniti2007,oldevik-dais2005,mwe2}.

AMMA (Atlas Model Management Architecture)~\cite{allilaire2006} is a model management platform with a component for global resource management, known as the AtlanMod MegaModel Management (AM3) tool.
AM3 supported automated megamodel discovery and was used for execution of compositions of ATL model transformations. Subsequently, other frameworks and tools for model management emerged, such as \cite{kolovos08epsilon, moscript2012, mdeforge2015, sandroSFKC15}. There exists some work in the MDE literature which apply megamodelling in specific domains~\cite{mgm1, simmonds2015}. ExecUtable RuntimE MegAmodels (EUREMA)~\cite{vogel_thesis} provides an MDE infrastructure for building self-adaptive software with the use of runtime megamodels. In order to manage the interconnection of the different models within the running system, EUREMA provides a runtime megamodel containing the related models as well as activities processing those models. This megamodel is automatically synchronized at runtime and executed to launch feedback loops linking design and runtime models. Similar to this work, our approach also supports dynamic update of the megamodel during enactment which will make it possible to link design and runtime models during NS deployment and management.

The MegaM@Rt2 ECSEL project~\cite{megamart} is working on a model-based framework for continuous development and runtime validation of complex systems. They use megamodels for traceability management with the goal of interrelating design and runtime models. It is not clear as yet what, if any, process modelling, enactment and transformation chaining means is covered.

A combination of transformation chaining and model management means has been used in the MDE community in different contexts with varying goals~\cite{fritzsche2011, uniti2007,  ftgpm-sdl2013}. The Epsilon framework~\cite{epsilon} supports orchestration of chains using Ant scripts\cite{ant} and also offers tools for model management.

The MDPE Workbench~\cite{fritzsche2011} is a framework for end-to-end performance decision support which uses model transformation chains for incorporating decision support in process modelling tool chains. The goal of this work is to create and analyze a process modelling tool chain by building a transformation chain of process models defined using BPMN, Process Flow, and JPass. 

UNiTi~\cite{uniti2007}, part of the MARTES project~\cite{martes}, is a model-based approach to construct, reuse and execute transformation chains in a technology independent way allowing loose coupling between implementation and specification. UNiTi focuses on data-flows and provides model management with the use of megamodels. In MAPLE, we go beyond that and cover both data and control flows explicitly in the process model.

The FTG+PM framework~\cite{ftgpm-sdl2013} which supports process execution based on an FTG (Formalism Transformation Graph - a subset of megamodels) is similar to our approach. Both data-flow and control-flow modelling is supported. However, the FTG and PM is defined as a single combined formalism, and so the FTG needs to be created together with the PM. There is no support for automated derivation or dynamic update of the MgM (which is a core part of our approach), and the research tool developed only supports execution of T-Core transformations. The FTG+PM, however, supports manual transformations during enactment, which we do not cover as yet.

Process enactment is a widely adopted method in the business process modelling domain. The model-based methods and tools in this domain mostly support SPEM/BPMN processes and use BPEL for orchestration~\cite{transforms-2009, aldazabal2008}. UML Activity Diagrams together with MDE enablers, such as model transformations, are not widely used in this area for modelling and enacting processes.
SPEM4MDE~\cite{diawLC11-spem4mde} is a process modelling language and environment for modelling MDE software processes. Process enactment is also supported based on UML state machines and QVT. ModelBus~\cite{aldazabal2008} supports definitions of tool chains and execution of software processes based on orchestration of transformation chains and process enactment. Model management is not available in these approaches.

MAPLE has been built on concepts taken from process modelling, process enactment, transformation chain orchestration, and model management with megamodelling. In essence, it is a megamodel-driven process enactment approach. It provides support for both control flow and data flow modelling and for execution of loosely coupled heterogeneous model transformation chains.  MAPLE offers direct support for ATL and Java transformations, and indirectly with the use of the \emph{executable process} module, it can support orchestration of chains implemented in a wide array of languages. The megamodel is dynamically updated during enactment which we believe makes it very useful for network service management. To the best of our knowledge, none of the projects, approaches, and tools covered in this section provide such support.

Our initial intention was to reuse and integrate existing components to build the MAPLE environment. However to the best of our knowledge, no working tools to serve our purpose exist at the moment. We looked into Eclipse-based tools including MoDISCO/AM3, UNiTi, TraCo for megamodelling support and Wires*, MWE2 (Modelling Workflow Engine), ATLFlow for orchestration support. None of the tools with megamodelling support were usable (incompatible with Eclipse Papyrus, unavailable, or failed to work) at the time of this work. Orchestration engines available were not adequate for our needs, since we wanted to support concurrent executions of model transformations. With regards to the translation engine, Wires* could have been adapted for our purpose since it supported orchestration of ATL transformation chains derived from UML Activity Diagrams. However, the tool is no longer maintained and not available for use. For this reason, we developed our own translation engine and orchestration engine as well as the underlying model management support which allowed us to offer a flexible and extensible integrated environment for process modelling and enactment in Papyrus.

In Table~\ref{tab:compTable}, we summarize and compare the related approaches using four criteria. The first criterion is about support for model management using megamodels. 
We further refine this criterion to see whether the approach supports automatic discovery of the resources and dynamic update of megamodels. The second criterion addresses two aspects of transformation chaining support: availability of automated support for chaining and execution, and model transformation languages supported. The approaches tagged as `generic' supports execution of heterogeneous chains. The third criterion is on process modelling support using UML Activity Diagrams or other notations. Finally, we have also investigated whether each approach supports process enactment.
As can be deduced from the comparison table, none of the related approaches and environments satisfies all the four criteria, while MAPLE has been designed to provide support for all of these features.

    \begin{table}
    \captionof{table}{Comparison of the approaches (supports (\cmark), does not support (\xmark) , unknown/unclear (\_))}
    \label{tab:compTable}
    \centering
    \tiny

\begin{tabularx}{\textwidth}{|p{.14\textwidth} | p{.04\textwidth} |  p{.07\textwidth} | p{.07\textwidth} | p{.05\textwidth} | p{.07\textwidth} | p{.08\textwidth} | p{.09\textwidth} | p{.06\textwidth} |}
\noalign{\hrule}
    \toprule
    \textbf{Approach}          & \multicolumn{3}{c |}{\textbf{Model Management}}                    & \multicolumn{2}{c |}{\textbf{MT Chaining}}                               & \multicolumn{2}{c |}{\textbf{Process Modelling}}         & \textbf{Process Enactment}     \\
   \noalign{\hrule}
                           & \textbf{MgM}  & \textbf{Auto. Discovery} & \textbf{Dynamic Update} & \textbf{Auto.} & \textbf{MT Language(s)}  & \textbf{UML AD } & \textbf{Other} &\\
                       & & &  &  &  &  & &       \\
\noalign{\hrule}
      \textbf{UNiTi~\cite{uniti2007}}             & \cmark  & \xmark & \xmark &  \cmark & ATL &  \cmark (only data flow) & \xmark & \_ \\
      \noalign{\hrule}
      \textbf{AM3~\cite{allilaire2006}}       & \cmark  & \_& \xmark &  \cmark & ATL & \xmark & \xmark & \xmark  \\
      \noalign{\hrule}
      \textbf{MegaM@Rt2~\cite{megamart}}     & \cmark &  \cmark &  \cmark & \xmark & \_  & \xmark & \_ &\_   \\
      \noalign{\hrule}
      \textbf{MCC~\cite{mcc2006}}     &\xmark  & \xmark & \xmark &  \cmark & generic & \xmark & \xmark & \_  \\
      \noalign{\hrule}
       \textbf{FTG+PM~\cite{ftgpm-sdl2013}}     & \cmark  & \xmark & \xmark &  \cmark & T-Core in AToMPM &  \cmark & \xmark &  \cmark  \\
       \noalign{\hrule}
       \textbf{EUREMA~\cite{vogel_thesis}}     & \cmark  &  \cmark &  \cmark & \xmark & \xmark & \xmark & \xmark & \xmark   \\
       \noalign{\hrule}
        \textbf{Epsilon~\cite{kolovos08epsilon}}     & \cmark  & \xmark & \xmark &  \cmark & Epsilon  & \xmark &  \cmark (only data flow) &\_   \\
        \noalign{\hrule}
         \textbf{MoScript~\cite{moscript2012}}     & \cmark  &\_ &  \cmark &  \cmark & ATL, QVT  & \xmark & \xmark & \xmark   \\
         \noalign{\hrule}
          \textbf{MDEForge~ \cite{mdeforge2015}}     & \cmark  & \_ & \_ & \xmark & \xmark  & \xmark & \xmark & \xmark   \\
          \noalign{\hrule}
          \textbf{MoTCoF~\cite{motcof2012}}     &\_  & \xmark & \xmark &  \cmark & generic & \xmark & \_ & \xmark   \\
          \noalign{\hrule}
           \textbf{Wires*~\cite{wires-mtatl2009}}     &\xmark  & \xmark & \xmark &  \cmark & ATL  & \xmark &  \cmark (only data flow) &  \cmark   \\
           \noalign{\hrule}
            \textbf{MMINT~\cite{sandroSFKC15}}     & \cmark  & \xmark &  \cmark & \xmark & \xmark  & \xmark & \xmark & \xmark   \\
            \noalign{\hrule}
             \textbf{Etien et al.~\cite{etien-amt2012}}     &\xmark  & \xmark & \xmark &  \cmark & QVTo  & \xmark & \xmark & \xmark    \\
             \noalign{\hrule}
             \textbf{Wagelaar~\cite{wagelaar2006}}     &\xmark  & \xmark & \xmark &  \cmark & generic   & \xmark &  \cmark (only data flow) & \xmark   \\
             \noalign{\hrule}
             \textbf{TraCo~\cite{traco}}     & \cmark  & \xmark & \xmark &  \cmark & ATL  & \xmark &  \cmark (only data flow) &  \cmark    \\
             \noalign{\hrule}
              \textbf{Aranega et al.~\cite{aranega-fmtc-2012}}     &\xmark  & \xmark &  \cmark &  \cmark & generic  & \xmark & \xmark & \xmark    \\
              \noalign{\hrule}
              \textbf{Juniper~\cite{juniper}}     &\xmark  & \xmark & \xmark &  \cmark & generic  & \xmark & \xmark & \xmark    \\
              \noalign{\hrule}
              \textbf{MWE 2~\cite{mwe2}}     & \xmark & \xmark & \xmark &  \cmark & MWE & \xmark & \xmark & \xmark   \\
              \noalign{\hrule}
              \textbf{Fritzsche et al.~\cite{mgm1}}     & \cmark  & \_ & \xmark &  \cmark & ATL &  \cmark & \cmark BPMN, JPASS, SAP proprietary languages & \cmark    \\
              \noalign{\hrule}
               \textbf{Simmonds et al.~\cite{simmonds2015}}     & \cmark  & \xmark & \xmark &  \cmark & ATL  &  \cmark & \cmark SPEM &  \cmark \\
               \noalign{\hrule}
               \textbf{Fritzsche et al.~\cite{fritzsche2011}}     & \cmark  &  \cmark & \xmark &  \cmark & ATL  &  \cmark & \cmark NetWeaver BPM  &  \cmark \\
               \noalign{\hrule}
                 \textbf{Maciel et al.~\cite{transforms-2009}}     &\xmark  & \xmark & \xmark &  \cmark & ATL & \xmark & \cmark SPEM 2 &  \cmark   \\
                 \noalign{\hrule}
                 \textbf{Aldazabal et al.~\cite{aldazabal2008}}     &\xmark  & \xmark & \xmark & \_ & \_  & \xmark &  \cmark SPEM, BPMN &  \cmark  \\
                 \noalign{\hrule}
                  \textbf{SPEMMDE~\cite{diawLC11-spem4mde}}     &\xmark  & \xmark & \xmark & \_ & QVT & \xmark & \cmark SPEM &  \cmark   \\
                  \noalign{\hrule}
     	         \textbf{Oldevik et al.~\cite{oldevik-dais2005}}     & \cmark & \xmark & \xmark &  \cmark & generic & \xmark & \xmark & \xmark   \\ \bottomrule
     	          \noalign{\hrule}
     	           \textbf{MAPLE}  & \cmark & \cmark & \cmark &  \cmark & generic & \cmark & \cmark & \cmark   \\ \bottomrule
     	          \noalign{\hrule}
    \end{tabularx}
    \end{table}
    \vspace{-5mm}

%% file: conclusion.tex
\label{sec:conclusion}

Automating NS management is one of the challenges in the NFV domain, and this is what we have aimed to address in this paper. This work resulted in a comprehensive and extensible environment, MAPLE, for model-driven process enactment. We have used the existing UML Activity Diagram environment in Papyrus and integrated process enactment means with it. 

In our approach, we followed the model-driven paradigm all through. The core of the approach combines orchestration of model transformation chains with model management means. We begin with a process which is modelled as a UML Activity Diagram (referred to as the PM). The activities in the PM are associated with model transformations. Input and output objects are model instances of some existing domain-specific language. For model management, we build a megamodel (MgM) of the target system. This MgM contains information of all MDE resources that are being used by the process, as well as the link(s) between these resources. The PM itself is also a resource which is registered in the MgM. To enact the PM, the MgM is used along with the PM to build a model transformation chain. Token-based enactment means have been implemented to orchestrate the MT chain. 

The enactment support has been created with NFV systems as the target domain. The approach along with the tool support is not restricted to NFV, and can be used in various domains for process enactment. We have demonstrated the use of MAPLE on a network service design and onboarding case study. The environment designed and implemented is not closed; it will be improved, extended further and validated.  